\documentclass[prd,aps,twocolumn,showpacs,preprintnumbers,amsmath,amssymb]{revtex4}
\usepackage{mathrsfs}
\usepackage{amsmath,amssymb,graphicx}
\usepackage {amssymb}
\usepackage{color}
\newcommand{\nc}{\newcommand}
\nc{\bea}{\begin{eqnarray}} \nc{\eea}{\end{eqnarray}}
\nc{\be}{\begin{equation}} \nc{\ee}{\end{equation}}

\input{epsf.sty}

\begin{document}


\title{Galileon Bouncing Inflation after BICEP2}

\author{Taotao Qiu}
\email{xsjqiu@gmail.com}
\affiliation{Institute of Astrophysics, Central China Normal University, Wuhan 430079, China}
\affiliation{State Key Laboratory of Theoretical Physics, Institute of Theoretical Physics, Chinese Academy of Sciences, Beijing 100190, China}


\begin{abstract}

We present a nonsingular scenario in which an inflation era goes after a bounce from a contracting scenario in the early universe. The contracting of the universe is supposed to be slow, such that the initial anisotropies will not grow too fast to become dominant and destroy the bounce. After the bounce, the universe enters into an inflationary region and reheating phase, where primordial perturbations are generated. The tensor-to-scalar ratio of the perturbations are expected to be consistent with the newly released data, $r=0.2^{+0.07}_{-0.05}$. The addition of the bounce process is aimed at getting rid of the annoying Big-Bang Singularity, which generally exist in pure inflation models.

\end{abstract}

\maketitle

\section{Introduction}
Recently many kinds of observational data in cosmology have been released fastly. After the announcement of WMAP9 \cite{Hinshaw:2012aka} and PLANCK data \cite{Ade:2013zuv} last year which greatly improved the accuracy of our measurement of the CMB sky, a few days ago the BICEP group in the South Pole has released their observation data of the BB-mode of the CMB polarization, claiming that they detected the gravitational waves directly \cite{Ade:2014xna}. There results shows that the ratio of tensor and scalar spectra of primordial perturbations (tensor-scalar-ratio), $r$, is lying within the region of \be r=0.2^{+0.07}_{-0.05}~,\ee while the $r=0$ point, indicating the absence of the gravitational waves, has been disfavored by $7\sigma$. This exciting results not only greatly supports the correctness of Einstein's gravity, but also, in cosmological sense, can distinguish various early universe models in a more confirming way. For example, it can strongly support inflation models which can generate large tensor perturbations, while disfavor models which cannot.

Although the observational results suggests an inflationary era in the early universe, however, theoretically speaking, traditional inflation scenario still needs to be questioned. One of the biggest problems that exist in inflation is the so-called singularity problem, which has been proved by Hawking et al. in their early work \cite{Hawking:1969sw,Borde:1993xh}. They claimed in their singularity theorem that the universe will meet the singularity if it satisfies the conditions {\it 1. the General Relativity holds}, and {\it 2. the Null Energy Condition (NEC) is preserved}. At the singularity, everything blows up and one can not get control of the universe under classical description. More seriously, Martin and Brandenberger claimed that in inflation models, if the number of efolds is longer than that is required to solve the Big-Bang problems, then the fluctuation modes will enter into a sub-Planckian zone of ignorance \cite{Brandenberger:2000wr}. Such problems occurs before the onset of inflation, therefore is difficult to be solved within inflation scenario itself. This motivates us to find alternative theories in pre-inflation era.

Phenomenologically, there might be a few evolutions that can be set in front of inflation in order to get rid of those problems. For example, the universe may undergo a contracting phase where the scale factor $a(t)$ shrinks initially and then, by some mechanism, ``bounces" into an expanding one \cite{Novello:2008ra}. The whole process can be done non-singularly if at the bouncing point $a(t)\neq 0$ \cite{Cai:2007qw}. The bouncing scenario has many interesting properties, for example, the Big-Bang puzzles such as horizon problem and flatness problem can be solved even in contracting phase, and scale-invariant primordial perturbations can be generated, etc \cite{Cai:2007zv}. Moreover, such non-singular scenario can also be non-trivially extended to the cyclic universe \cite{Xiong:2007cn,Piao:2004me}.

\section{bouncing inflation scenario}
In bouncing inflation scenario, one should first determine how fast the universe contracts, that is, how large the equation of state is in the contracting phase. In order for the universe goes successfully from contracting phase to bounce, the matter that drives the universe must be dominant during the contracting phase, which requires that its energy density of the matter grows fastest. However, this condition will be challenged if one takes into account of the initial anisotropy of the spacetime, which has been addressed as the ``anisotropy problem" \cite{Kunze:1999xp} of bouncing cosmology. To take a more clear example, one can start with a simple anisotropic Bianchi-IX metric \cite{Misner:1974qy}:
\be
ds^2=-dt^2+a^2(t)\sum_{i=1}^3e^{2\beta_i(t)}d{x^i}^2~,
\ee
with $\sum_{i=1}^3\beta_i(t)=0$. The anisotropy term will generate a energy density of anisotropy, $\Delta\rho_{ani}=(\sum_{i=1}^3\dot\beta_i^2)/2$. According to the Friedmann Equation, $\beta_i$'s evolve as:
\be
\ddot\beta_i+3H\dot\beta_i=0~,
\ee
which provide the solutions $\beta_i\propto a^{-3}(t)$ and induces that the anisotropy energy density increase as fast as $\Delta\rho_{ani}\propto
a^{-6}(t)$. Since in general the energy density of matter with equation of state $w$ grows as $a^{-3(1+w)}$, it corresponds to an effective energy density with EoS $w=1$, and will grow fast and become dominant over all species with EoS less than 1, leading finally to a collapsing anisotropic universe. This requires that, in order to have a successful bounce, the matter which drives the bounce must grow no slower than the anisotropy, requiring that its EoS no less than unity. Of course, in an expanding universe the anistropy decays fast as $a^{-6}$, thus isotropy can always be achieved, so we have nothing to worry about.

When the universe has a pre-inflationary contracting, primordial perturbations can be generated either in contracting phase, or in expanding phase. According to the perturbation theory, the equation of motion for the curvature perturbation is \cite{Mukhanov:1990me}: \be\label{eoms} u^{\prime\prime}+(c_s^2k^2-\frac{z^{\prime\prime}}{z})u=0~,\ee where $u\equiv z\zeta$, $z\equiv a\sqrt{2\epsilon}$ and $\zeta$ is the curvature perturbation, $c_s$ is the sound speed of perturbation, $\epsilon=3(1+w)/2$ is the slow-roll parameter, and prime denotes derivative with respect to conformal time, $\eta\equiv\int a^{-1}dt$. This equation can be solved to give the curvature perturbation: \be\label{sol} \zeta\sim k^{-\nu}~,~~~k^{\nu}|\eta_\ast-\eta|^{2\nu} \ee with $\nu=(\epsilon-3)/2(\epsilon-1)$ and $\eta_\ast$ is some integral constant. According to the definition of power spectrum: $P_\zeta\equiv k^3|\zeta|^2/(2\pi^2)$ and the spectral index: $n_s=1+d\ln P_\zeta/d\ln k$, one has $n_s=4-2|\nu|$.

There are two possibilities to get scale-invariant power spectrum, that is, to have $n_s\simeq 1$, as data suggests. One is $\epsilon\simeq 0(w\simeq-1)$, and in this case the constant mode should be dominant indicating that the varying mode is decaying, which can occur in expanding universe. This corresponds to the usually inflationary case. The other is $\epsilon\simeq3/2(w\simeq0)$, meaning that the varying mode is dominant, so this mode should be a growing one, which happens in contracting phase. This corresponds to the ``matter-like" contracting phase used in the so-called matter-bounce scenario \cite{Finelli:2001sr}. In other words, if the primordial perturbations are generated in contracting phase, it requires that the universe contracts with $w\simeq 0$, however contradicts with the constraints from the aforementioned anisotropy problem \footnote{There are actually several mechanisms to reconcile the inconsistency between scale-invariance and anisotropy. For example, in \cite{Cai:2012va} another period of contracting phase with $w>1$ has been added to the matter contracting phase, and in \cite{Qiu:2013eoa} a curvaton mechanism has been introduced to generate scale-invariant perturbations in an $w>1$ contracting phase. Here we focus on different possibility. }. In this case, we suggest that the perturbations (at least in the observable region) is generated in the inflationary era.

Moreover, according to the equation of motion for the tensor perturbations \be\label{eomt} v^{\prime\prime}+(k^2-\frac{a^{\prime\prime}}{a})v=0~\ee where $v\equiv ah$, similar results can be drawn for the gravitational waves $h$. The tensor spectrum and its index are defined as: $P_T\equiv k^3|h|^2/(2\pi^2)$ and $n_T=d\ln P_T/d\ln k=3-2|\nu|$, respectively.

\section{The Galileon bouncing inflation model}
In general, to get an inflation model, we only need a single scalar field with the lagrangian: ${\cal L}=K(X,\phi)$, where $X\equiv-\partial_\mu\phi\partial^\mu\phi/2$ is the kinetic term of $\phi$. However, if we also require the universe to realize bounce behavior, the single scalar field model cannot work and we need more degrees of freedom. As is well known, when the universe bounces from contracting phase ($H<0$) to expanding phase ($H>0$), the NEC will be violated \cite{Cai:2007qw}. The violation of NEC can be viewed as a critical point in breaking Hawking et al's Singularity Theorem to keep the singularity away, but in a usual field theory description, it will cause ``ghost" the quantization of which is unstable \cite{Carroll:2003st}. To eliminate the ghost, in this context we consider using the Galileon theory for the bounce realization \cite{Qiu:2011cy}.

We consider such a general form of lagrangian as:
\be\label{lagrangian} {\cal L}=K(X,\phi)-G(X,\phi)\Box\phi~,\ee where the last term is introduced by the Galileon theory \cite{Nicolis:2008in}, with $\Box\phi\equiv g^{\mu\nu}\nabla_\mu\nabla_\nu\phi$. The Galileon term shares the property that the equation of motion can remain second order due to the cancelation of the higher order terms, so it only brings a non-dynamical degree of freedom, which can help bounce the universe but without the ghost problem. The equation of motion of the Galileon field $\phi$ is \bea &&[K_X+2K_{XX}X-2(G_\phi+G_{X\phi}X)+6G_XH\dot\phi~\nonumber\\ &&+6HG_{XX}\dot\phi]\ddot\phi+3H[K_X-2(G_\phi-G_{X\phi}X)]\dot\phi~\nonumber\\ &&+[2K_{X\phi}+6G_X(\dot H+3H^2)-2G_{\phi\phi}]X-K_\phi=0~,\eea and the energy density and pressure are \be \rho=2K_XX-K+3G_XH\dot\phi^3-2G_\phi X~,~P=K-2(G_\phi+G_X\ddot\phi)X~.\ee

There are large possibilities of realizing bounce within this model, however, in this context, we will consider a rather simple picture, in which the Galileon term only affects on the bouncing point, and in the other regions, it will have negligible effects. Thus the function $G(X,\phi)$ will have a peak-shaped function located on the bouncing point. This can be realized if one writes $G(X,\phi)$ as $g(t)X$, where $g(t)$ has a peak at $t=t_B$. Assuming that the field $\phi$ runs monotonically with $t$, the function $g(t)$ can also be rewritten as a $\phi$-dependent function, $g(\phi(t))$. Therefore in regions far from bounce, the universe can be viewed as dominated by a single scalar field, $K(X,\phi)$.

In the contracting region where the universe started, $w>1$ is needed in order to eliminate the anisotropy dominance. We could start with an example where $K(X,\phi)$ takes a canonical form, namely $K(X,\phi)=X-V^{con}(\phi)$. Thus the energy density and pressure becomes \be \rho\simeq X+V^{con}~,~~~P\simeq X-V^{con}~,\ee and the EoS is simply $w=P/\rho=1-2V^{con}/\rho$. Since require the energy density of our model is positive definite, $w>1$ then requires a negative $V^{con}(\phi)$ in the contracting phase.

A negative $V^{con}(\phi)$ is not difficult to obtain, for example, if one takes the simple example of $V^{con}=-V_0 e^{c\phi}$ where $V_0$, $c$ are positive constants, one could get a nice attractor solution \cite{Qiu:2013eoa}. From the equation of motion \be \ddot\phi+3H\dot\phi+\frac{\partial V^{con}}{\partial\phi}=0~\ee one can get one of the solutions: \be
\phi(t)\simeq-\frac{2}{c}\ln(t_\ast-t)~,~~~\dot\phi(t)\simeq\frac{2}{c(t_\ast-t)}~,
\ee with the EoS $w\simeq c^2/3-1$, so it is rather easy to get a large $w$ provided that $c$ is large enough. From this solution one can also see that both $\phi$ and $\dot\phi$ goes monotonically with $t$. This is useful since as $t$ increases, the large velocity of the field will trigger the term of $G\Box\phi\sim G(\ddot\phi+3H\dot\phi)$, and bounce is possible to happen.

This model can be extended to more general cases. For example, one can have a non-canonical $X$-dependence such as $K(X,\phi)=k(\phi)X+t(\phi)X^2-V^{con}$, where $k(\phi)$ and $t(\phi)$ are arbitrary functions of $\phi$. We require that $k(\phi)\rightarrow 1$ and $t(\phi)\rightarrow 0$ in the far past, so it will hardly affect the previous results. However, both $k(\phi)$ and $t(\phi)$ can be important near the bounce. It can be shown that when we take the form of $k(\phi)$ where it can flip its sign at some pivot point, the velocity of $\phi$ may get greatly suppressed after the bounce, leading to a rather small $\dot\phi$ prepared for the entrance of inflation era \cite{Cai:2012va,Koehn:2013upa}.

In inflation era, the kinetic term of $\phi$ becomes very small, and we also require the Galileon term to be negligible such that $\phi$'s lagrangian is again dominated by $K(X,\phi)$. However, the negative potential cannot afford inflation, and we need a positive flat potential to have $w\simeq-1$. As an explicit example, here we take potential to be that of a small-field inflation potential, which is \be V^{inf}(\phi)=\Lambda^4(1-\frac{\phi^2}{\phi_0^2})^2~,\ee where $\Lambda$ is the height of the potential, while $\phi_0$ is the field value at which the potential is in its minimum, around which $\phi$ will oscillate and reheat at the end of inflation. This potential can be explained as originated from particle physics cosmology, such as Higgs-inflation \cite{CervantesCota:1995tz}. According to the potential, one can get the slow-roll parameter and the efolding number as: \bea \label{epsilon}\epsilon(\phi)\simeq\frac{M_p^2}{2}\left(\frac{V_\phi}{V}\right)^2=\frac{8M_p^2\phi^2}{(\phi^2-\phi_0^2)^2}~,\\ N\simeq\int^{\phi_e}_{\phi_i}\left(\frac{V}{V_\phi}\right)d\phi=\left(\frac{\phi^2}{8}-\frac{\phi_0^2}{4}\ln\phi\right)\bigg|^{\phi_e}_{\phi_i}~,\eea where $M_p$ is the Planck mass and $V_\phi=\partial V/\partial\phi$. Moreover, $\phi_i$ and $\phi_e$ indicate the initial and ending values of $\phi$ during inflation, respectively, and can be determined by requiring $\epsilon(\phi_e)=1$ and $N=60$. For example, a possible choice is made in which $\phi_0=13$ and $M_p=1$ with arbitrary $\Lambda$ \cite{Ade:2013zuv}, one has $\phi_i\approx 1.9$ and $\phi_e\approx 11.66$. Note that in inflationary era, the field $\phi$ runs also monotonically with $t$.

One can also generalize the inflationary side of the model by introducing prefactor functions such as $k(\phi)$ and $t(\phi)$, and requiring that they have trivial effects during inflationary era. Moreover, the potential in contracting and expanding periods can be phenomenologically glued together to describe the whole theory. One example is \be V(\phi)=[1-\tanh(\lambda_1\frac{\phi}{\phi_B})]V^{con}(\phi)+[1+\tanh(\lambda_2\frac{\phi}{\phi_B})]V^{inf}(\phi)~,\ee with $\lambda_1,\lambda_2\ll 1$ and $\phi_B$ is the value of $\phi$ at bouncing point. This can be done thanks to the monotonicity of $\phi$ with respect to $t$.

Note that if the equation of state in contracting phase is $w=1$, the model can minimally satisfy the requirement of solving the anisotropy problem. This case can be realized by letting the kinetic term of the model much larger than the potential, but the potential is not necessarily negative. A positive potential seems even more natural to connect with an inflationary potential. This kind of model has been studied in e.g. \cite{Piao:2003zm}.

Let's now focus on the perturbations generated in this model. After taking the uniform-$\phi$ gauge, one can perturb the lagrangian (\ref{lagrangian}) up to the second order to obtain the perturbed action: \be
{\cal S}^{(2)}=\int d\eta d^3xa^2\frac{Q}{c_{s}^2}\Bigl[\zeta^{\prime 2}-c_{s}^2(\partial\zeta)^2\Bigr]~,
\ee where \bea Q&=&2M_p^4X[K_X-2(G_\phi-G_{X\phi}X)+2(G_X+G_{XX}X)\ddot\phi\nonumber\\ &&+4HG_X\dot\phi-2G_X^2X^2/M_p^2]/(M_p^2H-G_XX\dot\phi)^2~,\\ c_s^2&=&\frac{(M_p^2H-G_XX\dot\phi)^2}{2M_p^4X}\{K_X+2K_{XX}X-2(G_\phi+G_{X\phi}X)\nonumber\\ &&+6H(G_X+G_{XX}X)\dot\phi+6G_X^2X^2/M_p^2\}^{-1}Q~,\eea respectively. In most of the region in our model where $\phi$ has the canonical approximation, one have $Q\simeq 1$ and $c_s^2\simeq 1$. The equation of motion for $\zeta$ then reduce to (\ref{eoms}). Moreover, since there is no nonminimal coupling between $\phi$ field and gravity, the equation of motion for gravitational waves remains unaffected, which is (\ref{eomt}).

From the above background analysis, one can plot the Hubble horizon of our model as in Fig. 1.

\begin{figure}[htbp]
\centering
\includegraphics[scale=0.3]{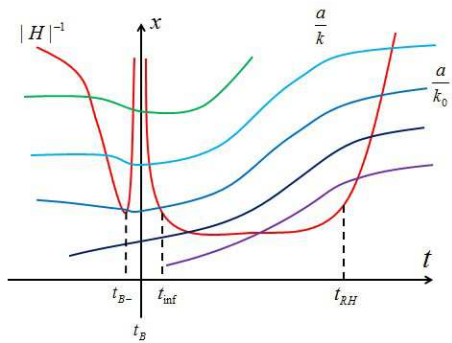}
\caption{The sketch plot of the horizon of our model (red) and the fluctuation modes that cross the horizon before or during inflation period (from green to purple). Since before the bounce the universe evolves slowly with a small value of Hubble parameter, the Hubble horizon is large at most of the time, with a sharp damp near the bounce. However, when bounce happens, it goes to infinity as $H=0$. At the inflation era when $H$ becomes nearly a constant, the horizon also behaves as a constant. The fluctuations with $k\leq k_0$ exit the horizon at contracting phase, thus getting a blue spectrum due to the nontrivial background. Among those the $k_0$ mode is the last mode, which corresponds to the bouncing energy scale. Those with $k>k_0$ exits the horizon at inflationary era, therefore obtaining a scale-invariant spectrum consistent with the observational data.}\label{horizon}
\end{figure}

From the plot one can see that, all the fluctuation modes which are generated inside the Hubble horizon has been divided into two parts, by some pivot scale $k_0$. The modes with $k<k_0$ are generated in the contracting phase, and will exit the horizon before bouncing, while the modes with $k>k_0$ are generated in expanding phase, and will exit the horizon after bouncing. Due to such difference, the super-horizon behavior of those two types of modes behaves totally different. Note that in pure bounce or inflation models, all the fluctuation modes are generated and exit horizon within the same background, and will have the same super-horizon behavior.

We assume that all the fluctuation modes are generated in Bunch-Davies vacuum, making $\zeta_i\sim e^{ikx}/\sqrt{2k}$. For $k<k_0$ modes, we can use equation of motion (\ref{eoms}) with $\epsilon\geq 1$ for large $w$. According to the Friedmann equation, one could parameterize scale factor $a$ in terms of $\eta$ as: $a(\eta)\sim|\eta_\ast-\eta|^{1/(\epsilon-1)}$ for arbitrary constant $\epsilon$. From this one can see, $|\eta_\ast-\eta|$ decreases as $a(\eta)$ shrinks. Moreover, $\nu=(\epsilon-3)/2(\epsilon-1)$ is a positive value, so the varying mode in solution (\ref{sol}) is actually a decaying one, and the power spectrum and the spectral index will be determined by the constant modes of $\zeta$. By rigidly solving the equation (\ref{eoms}) and according to the definition of $P_\zeta$ and $n_s$, one can finally obtain that the power spectrum and the spectral index for scalar perturbations of $k<k_0$ are: \be P^{con}_\zeta=\frac{H_{B-}^2}{8\pi^2M_p^2\epsilon}\left(\frac{k}{k_{B-}}\right)^{n^{con}_s-1}~, n_s^{con}=1+\frac{2\epsilon}{\epsilon-1}~,\ee and from the minimal requirement of solving anisotropy problem $w=1(\epsilon=3)$ to $w\gg 1(\epsilon\gg3)$, one has $3\leq n_s^{con}\leq 4$, which has a strong blue tilt. The same results holds for tensor perturbations. According to the equation (\ref{eomt}), the tensor modes generated in contracting phase also blue tilted, with the spectrum and the index as $P^{con}_T=2H_{B-}^2/(\pi^2M_p^2)(k/k_{B-})^{n^{con}_T}~$, $2\leq n^{con}_T=2\epsilon/(\epsilon-1)\leq 3$ for $3\leq \epsilon\leq \infty$. Finally, the tensor-scalar-ratio of those modes can be expressed as: \be r^{con}=\frac{P^{con}_T}{P^{con}_\zeta}=16\epsilon~.\ee

Similar analysis can be done on the fluctuation generated in expanding phase. Since in inflationary era $\epsilon\simeq 0$, as $a(\eta)$ expands $|\eta_\ast-\eta|$ still decreases, and since now $\nu$ is still positive, the varying modes in solution (\ref{sol}) is also a decaying one. The power spectrum and its index will be determined by the constant mode of $\zeta$. Rigid calculation gives: \be P^{inf}_\zeta=\frac{H^2}{8\pi^2M_p^2\epsilon}~, n_s^{inf}=1+\frac{2\epsilon}{\epsilon-1}\simeq1-2\epsilon~,\ee which is nearly scale-invariant. Similarly, the tensor perturbations also behave as scale-invariant, $P^{inf}_T=2H^2/(\pi^2M_p^2)$, $n^{inf}_T=2\epsilon/(\epsilon-1)\simeq-2\epsilon$. The tensor-scalar-ratio of those modes is then obtained as: \be r^{inf}=\frac{P^{inf}_T}{P^{inf}_\zeta}=16\epsilon~.\ee

\begin{figure}[htbp]
\centering
\includegraphics[scale=0.3]{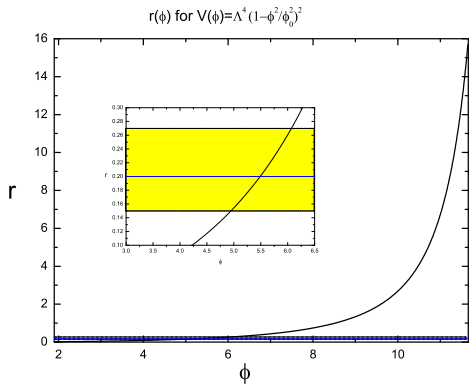}
\caption{The tensor-scalar-ratio $r$ obtained during inflationary era. The initial and final value of $\phi$ are set as $\phi_i=1.9$ and $\phi_e=11.66$ as obtained in the above context. The region allowed by BICEP2 is labeled as yellow. }\label{horizon}
\end{figure}

From the above we can see that, if the perturbation modes with large wavenumber $k>k_0$ enters into horizon today, we will see an scale-invariant scalar and tensor spectrum, which is consistent with the observational data \cite{Ade:2013zuv}. The tensor-scalar-ratio is proportional to slow-roll parameter $\epsilon$, which has been given in (\ref{epsilon}), and is plotted in Fig. 2. From the plot we can see, if the fluctuation mode that exit the horizon at the point where $r$ is within the allowed region is about $0.002\text{Mpc}^{-1}$, then the model is consistent with the BICEP2 data \cite{Ade:2014xna} \protect\footnote{There are large possibilities that it can be satisfied. For example, it is shown in \cite{Kobayashi:2010cm} that $r\simeq0.17$ can be given in G-inflation models. Moreover, the potential given in this context can also give rise to $r$ that is consistent with the data \cite{Freese:2014nla}.}. Moreover, the modes with smaller wavenumber $k<k_0$ hasn't enter the horizon yet, so its prediction of blue-tilted spectrum will be tested in future observations. However, actually we have already got some hints in favor of such a $k$-dependence, coming from the TT spectrum of CMB which is suppressed in small $l$ region ($l<10$) \cite{Ade:2013zuv}. Since the TT spectrum corresponds to the scalar perturbations, its suppression may be explained as the blue-tilt of the perturbations in pre-inflation evolution \cite{Piao:2003zm}. According to the phenomenon, the pivot scale $k_0\sim k(l\simeq 10)\simeq 0.001\text{Mpc}^{-1}$, and since the fluctuation modes with wavenumber $k_0$ can be viewed as the last mode that exits horizon during contracting phase, it corresponds to the comoving Hubble scale of bouncing, $a_{B-}H_{B-}$. This could furtherly constrain the energy scale of the bounce \cite{Cai:2007zv}.

\section{Final Remarks}
Recent observational data has discovered primordial gravitational waves with a non-trivial tensor-scalar-ratio, $r=0.2^{+0.07}_{-0.05}$, supporting the inflation scenario in the early universe. However, theoretically speaking inflation has been suffering from singularity and trans-Planckian problems. In this letter, we present a scenario where inflation is preceded by a non-singular bounce, solving the theoretical problems and meeting with the observational data simultaneously.

In our scenario, our universe starts with a slowly-contracting phase with EoS $w\geq 1$, such that the initial anisotropies, if exist, will not go too fast so as to dominate over the background, and make the universe collapse into a totally anisotropic one. This generally requires a negative potential of the cosmic field in the contracting phase (or kinetic term much larger than potential in $w=1$ case). After contracting, the universe bounces into an expanding phase with a positive flat potential, driving ordinary inflation. The bounce can be realized by introducing the ``Galileon-term" in the field lagrangian, such that when bounce violates NEC, ghost problem will not appear.

Primordial perturbations can be generated either in contracting phase, or in expanding phase. Those generated in expanding phase corresponds to large wavenumber $k$, which has entered into horizon and can be observed by nowadays observations. Taking proper inflation potential, one can have scale-invariant $n_s$ as well as large $r$, which is consistent with the BICEP2 data. Meanwhile, those generated in contracting phase corresponds to small $k$, which is favored by the small $l$ suppression observed at TT spectra of CMB map, and large range of blue tilt can be predicted for future observations. Moreover, in order for the pivot scale $k_0$ to be consistent with CMB, the energy scale of bounce can be constrained. We will leave these subjects for a future discussion \cite{qiutt}.

\begin{acknowledgments}
The author thanks Yun-Song Piao for his useful comments. This work is supported by the Open Project Program of State Key Laboratory of Theoretical Physics, Institute of Theoretical Physics, Chinese Academy of Sciences, China (No.Y4KF131CJ1).
\end{acknowledgments}


\end{document}